# ATTENUATION OF SHORT STRONGLY NONLINEAR STRESS PULSES IN DISSIPATIVE GRANULAR CHAINS

S. Y. Wang[1], V. F. Nesterenko[1,2*]

[1]Materials Science and Engineering Program, University of California at San Diego, La Jolla CA 92093-0418 USA

[2]Mechanical and Aerospace Engineering Department, University of California at San Diego, La Jolla CA 92093-0411 USA

PACS numbers: 05.45.Yv, 45.70.-n, 46.40.-f

Key words: granular chain, strongly nonlinear, dissipation, impact mitigation

**Abstract.** Attenuation of short, strongly nonlinear stress pulses in chains of spheres and cylinders was investigated experimentally and numerically for two ratios of their masses keeping their contacts identical. The chain with mass ratio 0.98 supports solitary waves and another one (with mass ratio 0.55) supports nonstationary pulses which preserve their identity only on relatively short distances, but attenuate on longer distances because of radiation of small amplitude tails generated by oscillating small mass particles. Pulse attenuation in experiments in the chain with mass ratio 0.55 was faster at the same number of the particles from the entrance than in the chain with mass ratio 0.98. It is in quantitative agreement with results of numerical calculations with effective damping coefficient 6 kg/s. This level of damping was critical for eliminating the gap openings between particles in the system with mass ratio 0.55 present at lower or no damping. However with increase of dissipation numerical results show that the chain with mass ratio 0.98 provides faster attenuation than chain with mass ratio 0.55 due to the fact that the former system supports the narrower pulse with the larger difference between velocities of neighboring particles. The investigated chains demonstrated different wave structure at zero dissipation and at intermediate damping coefficients and the similar behavior at large damping.

## I. INTRODUCTION

Granular chain with strongly nonlinear interaction between elastic spherical particles [1] of the same mass presents an example of the discrete systems with qualitatively new wave dynamics different than behavior of materials with elastically linear or weakly nonlinear interactions between elements [2-12]. One dimensional granular chains not precompressed by a static force represent an example of "sonic vacuum" [5]. This name emphasizes that sound speed in its classical sense in such system is zero. "Sonic vacuum" or weakly precompressed granular chains (dynamic strains being much larger than initial strains) supports strongly nonlinear waves which are qualitatively different than weakly nonlinear solitary waves [2-12]. For example, strongly nonlinear solitary wave has a width which does not depend on its amplitude, unlike width of weakly nonlinear solitary wave. The number of particles in this solitary wave depends on the details of interaction force, for example on the exponent in power law for interaction force [5]. Width of solitary wave, composed mostly from five particles in chains of spherical beads, comes from force exponent 3/2 in Hertzian interaction force. Ratio of solitary wave speed in sonic vacuum to initial sound speed is infinity, while for weakly nonlinear solitary wave this ratio is close to one. Strongly nonlinear solitary wave can be considered as "quasiparticle" with effective mass ~1.4 mass of the elements in Hertzian chain. A strongly nonlinear solitary wave converged into weakly nonlinear when strain amplitude is close to the initial value, the opposite transition is not valid [5].

The behavior of these waves was investigated by different groups of researchers numerically, for example [2,6,7], and experimentally using gauges embedded in the particles made from different materials [3,4,5, 8-11] and also high speed photography allowing to measure particles displacements with a micrometer-scale resolution [12]. Theoretical studies include exact analytical solution of strongly nonlinear wave equation [2, 5], stability of periodic solution in "sonic vacuum" [5,13], proofs of existence of solitary wave in discreet systems [14,15] and single pulse character with double exponential decay in [16-19].

Two mass, strongly nonlinear chains (dimer systems consisting of alternatively arranged particles with two different masses) demonstrate a new very interesting behavior which was first presented in [5] for relatively short chain composed from 40 particles. The long wave approximation in an extreme case, when the mass of one particle ($m_1$) is much larger than the mass of another ($m_2$) ($k = m_1/m_2 >> 1$), and both have the same diameter $a$, the system supports solitary waves with a characteristic space scale $L \approx 10a$ being twice that in the case of particles with the same masses.

Behavior of two mass chains with different mass ratios (2, 4, 16, 24, 64) have been studied numerically keeping the chain's macroproperties (linear density and elastic properties) the same, and equal to the properties of the chain with equal mass $m$. It was ensured by redistribution of masses between neighboring particles under the condition $2m = m_1 + m_2$ and choosing the same distances between particle centers and interaction constants for both types of chains. The force acting on the contact of the chain with a supporting wall was used for the comparison of pulse attenuation in different systems [5].

Numerical simulations demonstrated that initial disturbance (created by impact of particle with mass equal to 2.5 mass of the cell ($5m$)) in the chain with large mass ratio (64) was quickly transformed into three solitary-like waves, unlike in the chain with equal mass particles where six clearly detectable solitons were created. This behavior is in agreement with the discussed analytical prediction for $k \gg 1$. The phase speeds of leading pulses in both systems are practically identical. The time history of the light particle was similar to the velocity profile for the neighboring heavy particle only with amplitudes much smaller than amplitude of the leading pulse for the former particle. Additionally the motion of light particles exhibits a qualitatively new feature: high-frequency small-amplitude modulation of the velocity profile, more pronounced with the decrease of wave amplitude.

A new type of behavior was observed for a chain with smaller mass ratio despite keeping the same global properties of the system and impact by the same striker with the same velocity. When the mass ratio for the particles in the chain become equal to 2 or 4, there is no solitary waves formed based on the observation of the velocity history of light and heavy particles and force acting on the supporting wall. In this case the light and heavy particles behave significantly differently (see Fig. 1.18(b) and (c) in [5]). The velocity of the light particles in leading pulse has two maximums following by oscillatory tail with amplitude about two times smaller than amplitude of the leading pulse, even negative velocities were present in this oscillatory tail. These stress waves with complex shape mostly preserve their shape and amplitudes as they propagate along the investigated chain with 80 particles. Therefore, on this space scale they may be characterized as quasitationary waves.

The velocity of the heavy particles in leading pulse is represented by an asymmetric single peak followed by another peak with a wider duration and smaller amplitude. These two peaks mostly preserve their shape and amplitudes as they propagate with different speeds along the investigated chain with 80 particles, with an increase of space between them. These results were later confirmed in experiments.

It is important that at the same macroproperties and at the same striker impact, the systems with different mass ratios demonstrate better mitigation properties (charged by force acting on the supported wall) in comparison with the system with equal masses of particle. The difference is more than twice at optimal mass ratio demonstrating the possibility of optimization of the granular chains as nondissipative impact protectors keeping their global properties the same [5].

Very important new type of behavior of stress pulses in strongly nonlinear dimer chains excited by $\delta$-force applied to the first particle (global properties of these chains were not kept constant) was observed numerically in [20]. At certain discrete mass ratios of light to heavy spherical particles ($\varepsilon_n$ = 0.3428, 0.1548, 0.0901 and also at other smaller values of $\varepsilon_n$) a true solitary wave was observed in numerical calculations. This solitary wave propagated without any detectable attenuation over long distances in system with a total number of 251 beads. Its behavior is explained by the antiresonances in the dimer chain satisfied only for certain values of $\varepsilon_n$. These specific quantum values of mass ratios ensure unique behavior of particles in the wave - the synchronization of the motion of light and heavy beads, providing transferring of the entire energy of the pulse through the chain. The conditions of propagating of unattenuated compression pulse were also formulated using the asymptotic analysis based on slow-fast time scale separation of the system dynamics in a reasonable agreement with numerical simulation of discrete system [20].

At general values of $\varepsilon$ localized soliton-like propagating stress pulses were also observed, but their amplitude slowly decayed with distance and they were accompanied by oscillating tails. At values of $\varepsilon$ different from specific quantum values $\varepsilon_n$ the light bead loses contact with its left heavy neighbor at the end of the compression pulse, generating oscillations. They left behind the main compression pulse taking away its energy and resulting in its decay. In numerical calculations the maximum attenuation of compression stress pulse was observed at value $\varepsilon=0.59$ [20].

This behavior was confirmed in recent experiments performed by Potekin et al in a chain of 21 spheres suspended on the rods to minimize the dissipation effects [21]. Three chains were investigated with values of $\varepsilon=1$ (homogeneous chain), and two dimer chains with $\varepsilon=0.5$, and 0.125. The homogeneous chain and dimer chain ($\varepsilon=0.125$) supported solitary wave. The dimer chain with $\varepsilon=0.5$ demonstrated stronger attenuated behavior of the main pulse followed by oscillating tail. Three different levels of impacting forces were used to excite single pulses at the impacted end of the chain. The experimental results agree with numerical calculations (using damping coefficient in the range 32-35.4 Ns/m) demonstrating an expected deep minimum in transmitted force nearby $\varepsilon=0.5$, characteristic for nondissipative chain.

Dissipation is present in all experiments with granular chains. This dissipation can be due to viscoelastoplastic deformation of contacts [22,23], which in general has nonlinear dependence on strains [24-26]. Another strongly nonlinear discrete metamaterial composed from steel cylinders and rubber o-rings (with better tunability than system with Hertzian contacts [27-30]), demonstrated nonlinear dependence on strains and strain rates being sensitive to loading path [31,32].

Simpler models were also used to account for the dissipative properties of contact interaction using approach based on coefficients of restitution [33], viscous friction [34] or using standard viscous dissipation model depending only on strain rate [21,35-36]. These approaches allowed analysis of the unique role of dissipation on the pulse nature. For example, at certain dissipation level excitation by $\delta$-force resulted in two wave structure [35,36]. This dissipation model allowed to establish analytical conditions for the transition from oscillatory to monotonous shock wave profiles [37,38]. In this case the damping coefficient is a new effective parameter which may account for the complex nature of the dissipative processes during contact interaction. Of course its validity needs to be checked experimentally and may depend on conditions of experiments and material properties.

In this paper we conducted experiments and numerical calculations to check if two mass chains with mass ratio (0.55), being close to optimal mass ratio for attenuation in nondissipative chain, is still the preferable attenuating system also in the presence of dissipation. To keep mechanism of dissipation identical for both systems we used steel cylinders in contact with spheres. The mass ratio was changed only by the changing the height of cylinders keeping spheres the same thus preserving the type of contact interaction between neighboring particles unlike in case where diameters of spheres were changed [20,21]. This allowed to clarify the role of nonlinear dispersion caused by periodic arrangement of particles on pulse attenuation. The comparison of the attenuating properties of these two systems were made at the same number of particles from impacted end and also at the same mass of the system above particles where impulse was detected.

In the investigated range of pulse amplitudes, the linear viscous model with damping coefficient 6 kg/s satisfactorily described not only the attenuation of pulse amplitudes, but also the transformation of their shapes in both systems. The value of damping coefficient was significantly smaller than 32-35 kg/s in [21] probably due to the different nature of contacts. Two mass chain with mass ratio 0.55 demonstrated better performance in experiments and in numerical calculations only when the damping coefficient was below critical value. In numerical calculations with increased dissipation above this value, the one mass chain outperfoms the two mass system in attenuating pulse amplitudes.

## II. EXPERIMENTAL SET UP

Experiments were conducted with chains composed from SS304 stainless steel cylinders and 440C stainless steel balls (Fig. 1) arranged alternatively. Stainless steels SS304 and 440C have a similar elastic properties and density. The chain was placed vertically inside the channel made by four Al alloy rods providing alignment of particles with minimum contacts between them and rods for minimizing friction and dissipation. Two different mass ratios were achieved by keeping the spheres the same, while selecting cylinders with different heights. The larger cylinder has a height $h$=9.6 mm and a mass ($m_c$) of 3.77 g and the small cylinder has a height $h$=5.3 mm and a mass of 2.043 g. In both cases the spheres had a mass ($m_s$) 2.085 g and diameter $d$=8 mm, the same as diameter of cylinders. The cell size in this array was equal to $h+d$. The mass ratio of the chain with the smaller cylinder is 0.98 corresponding to the chain which supported solitary wave. For the system with the larger cylinder height, the mass ratio is 0.55, which corresponds to the most efficient attenuation due to dispersion effects in nondissipative chain of spheres [20]. The waves in the systems were generated by impact of 440C steel ball with mass 2.085 g (less than the cell mass equal 5.855 g and 4.128 g).

Two piezogauges were embedded inside cylinders at different depths, with one gauge embedded in 4$^{th}$ cylinder (used to specify the incoming pulse) and the other one inside the cylinder placed at different depths. The piezogauges, supplied by Piezo Systems Inc. were custom cut and wired, they had the sensitivity in the range 6.8 - 7.1 N/V. The piezogauges (RC~537 µs), were calibrated using the impact by PTFE sphere (mass 0.12 g) with a recorded velocities (0.7 – 0.8 m/s) and based on linear momentum conservation which gives the value of the force integral over time from the start of the impact to the maximum force. The signals from the gauges were recorded using oscilloscope Tektronics TDS 2014.

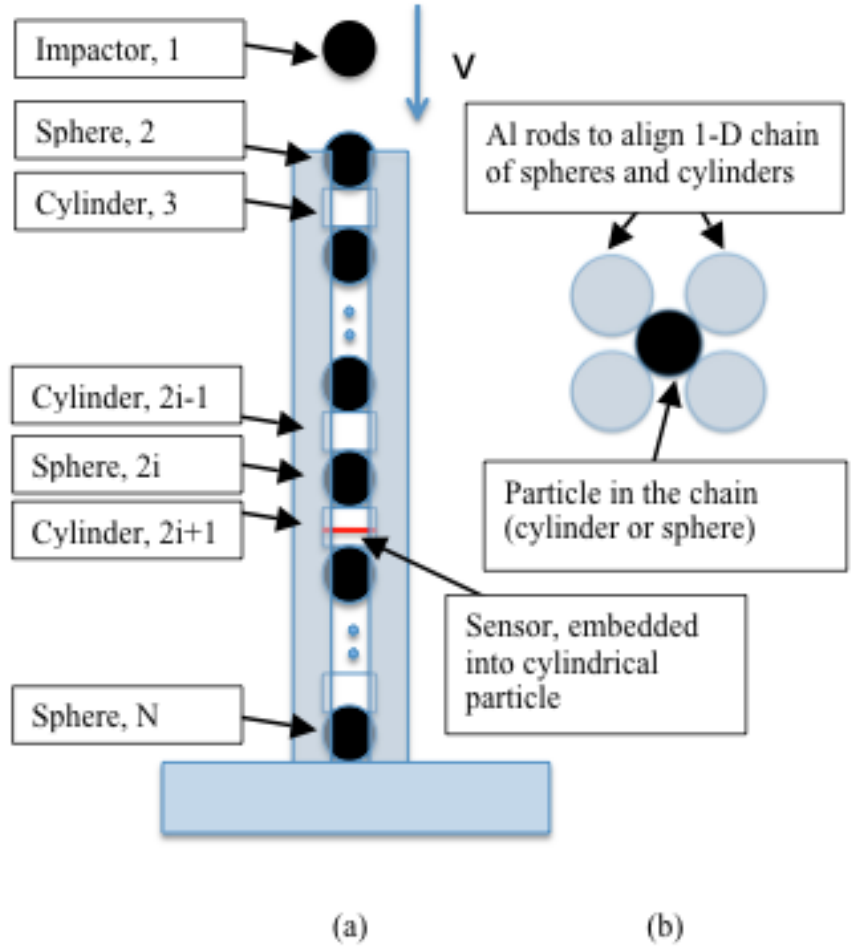

FIG. 1. (Color online) Experimental set-up. Cylinders and spheres aligned in 1-D chain inside the holder (a) and cross-sectional view of the assembly (b). Four aluminum rods hold the particles in aligned chain. In the numerical calculations the particles inside the chain with even numbers are spheres and particles with odd numbers are cylinders, $i = 2,3,4,…N/2$. The impactor is a separate particle outside of chain with number 1.

## III. NUMERICAL CALCULATIONS

Numerical calculations of the chain behavior were based on strongly nonlinear contact force between bodies with the radii of contact equal $R_1$ and $R_2$ described by a static Hertz law [1] valid for elastic contact deformation

$$F = \frac{4E_1 E_2}{3\left[E_1(1-\nu_2^2) + E_2(1-\nu_1^2)\right]} \left(\frac{R_1 R_2}{R_1 + R_2}\right)^{1/2} \delta^{3/2}, \qquad (1)$$

where $R_1$ and $R_2$ are corresponding radii of the contacting undeformed particles and their Young's moduli ($E_1$, $E_2$) and Poisson's ratios ($\nu_1$, $\nu_2$) and $\delta$ is the change between centers of interacting bodies due to contact deformation. The curvatures of the surfaces of contacting neighboring particles can be selected independently of their masses. These arrays may have large masses of cylinders with small radii of rounded ends. It allows change of interaction law and the properties of the system, but keeping overall density the same.

To apply static law (Eq. 1) for the dynamic contact interaction between particles the pulse duration should be much longer than the characteristic times of sound propagation in the sphere and cylinders equal to their characteristic sizes (diameters, height divided by corresponding sound speeds). The nondissipative equations of motion for particles inside the chain oriented in the vertical direction and thus including gravitational force are the following

$$m_c \ddot{u}_{2i-1} = A\left[\left\{(\upsilon_{2i-2,0} + \upsilon_{2i-2}) - (u_{2i-1,0} + u_{2i-1})\right\}_+^{3/2} - \left\{(u_{2i-1,0} + u_{2i-1}) - (\upsilon_{2i,0} + \upsilon_{2i})\right\}_+^{3/2}\right] + m_c g, \quad (2)$$

$$m_s \ddot{\upsilon}_{2i} = A\left[\left\{(u_{2i-1,0} + u_{2i-1}) - (\upsilon_{2i,0} + \upsilon_{2i})\right\}_+^{3/2} - \left\{(\upsilon_{2i,0} + \upsilon_{2i}) - (u_{2i+1,0} + u_{2i+1})\right\}_+^{3/2}\right] + m_s g, \quad (3)$$

where $m_c$ is the mass of the stainless steel cylinder and $m_s$ is the mass of the stainless steel sphere. Displacements $u_{2i-1,0}$ and $\upsilon_{2i,0}$ represent equilibrium displacements of centers of cylinders and spheres in the gravitationally loaded chain calculated from positions in initially undeformed chain. The other displacements $u_{2i-1}$, $\upsilon_{2i}$ represent dynamic parts of overall displacements during wave propagation. $N$ is the total number of particles (including the impactor), $i = 2,3,4,…N/2$, and even number $N$ corresponds to the last spherical particle contacting the wall. Even particle numbers correspond to spherical particles in the chain and odd numbers to cylinders, the spherical impactor is particle number 1 (see numbering of particles in Fig. 1). Positive subscript corresponds to the force between neighboring particles being in contact, otherwise interaction force is zero.

Coefficient $A$ for contact interaction of flat surfaces of cylinders and spheres with radius $R_S$ is equal

$$A = \frac{4 E_C E_S (R_S)^{1/2}}{3\left[E_S(1-v_C^2) + E_C(1-v_S^2)\right]}. \tag{4}$$

The constant $A$ depends on the Young moduli ($E_c$, $E_s$) and Poisson's ratios ($v_c, v_s$) of materials of interacting particles and the radius of sphere [1].

The separate equation for the impactor (dynamic displacement $\upsilon_1$), initially contacting the first sphere in the gravitationally loaded chain (there is no contact deformation between these two particles prior to the impact) is:

$$m_{imp}\ddot{\upsilon}_1 = -A_1(\upsilon_1 - \upsilon_2)_+^{3/2} + m_{imp}g, \quad A_1 = \frac{4 E_{imp} E_S (1/R_S + 1/R_{imp})^{-1/2}}{3\left[E_S(1-v_{imp}^2) + E_{imp}(1-v_S^2)\right]} \tag{5}$$

where $A_1$ is corresponding to the contact of the impactor and the first sphere having the same radii and elastic properties.

Equation for the first spherical particle in the chain (dynamic displacement $\upsilon_2$) is

$$m_s\ddot{\upsilon}_2 = A_1(\upsilon_1 - \upsilon_2)_+^{3/2} - A\left((\upsilon_{2,0} + \upsilon_2) - (u_{3,0} + u_3)\right)_+^{3/2} + m_s g. \tag{6}$$

Equation for the last spherical particle (dynamic displacement $\upsilon_N$) contacting the flat wall is

$$m_s \ddot{\upsilon}_N = A\big((u_{N-1,0} + u_{N-1}) - (\upsilon_{N,0} + \upsilon_N)\big)_+^{3/2} - A(\upsilon_{N,0} + \upsilon_N)_+^{3/2} + m_s g. \tag{7}$$

In all experiments we observed attenuation of the pulse amplitude. To explain this phenomenon we added linear viscous term ($F_{vis}$) to all contact interactions. Introduction of effective viscosity to describe dissipation processes (friction, viscoplastic deformation) on the contacts is similar to the one used in [21, 34-38]. Resulting viscous forces acting on the impactor, first particle, on cylinders and spheres inside the chain and between last particle and the flat wall at the deformed contacts are described by Eqs. (8)-(12) with corresponding coefficients of viscous damping $\mu_1$ and $\mu$,

$$F_{vis,1} = \mu_1 [\dot{\upsilon}_2 - \dot{\upsilon}_1], \tag{8}$$

$$F_{vis,2} = \mu_1 [\dot{\upsilon}_1 - \dot{\upsilon}_2] + \mu [\dot{u}_3 - \dot{\upsilon}_2], \tag{9}$$

$$F_{vis,2i-1} = \mu [\dot{\upsilon}_{2i-2} - 2\dot{u}_{2i-1} + \dot{\upsilon}_{2i}], \tag{10}$$

$$F_{vis,2i} = \mu [\dot{u}_{2i-1} - 2\dot{\upsilon}_{2i} + \dot{u}_{2i+1}], \tag{11}$$

$$F_{vis,N} = \mu [\dot{u}_{N-1} - 2\dot{\upsilon}_N]. \tag{12}$$

When the particles are separated the viscous term is considered to be equal zero. In numerical calculations of dissipative chain a linear momentum were conserved with accuracy $10^{-6}$ %.

In the numerical calculation the pulse was generated by giving an initial velocity to the impactor. Its velocity was adjusted to provide the amplitude of the reference pulse similar to the corresponding experimental values for adequate comparison of its evolution at later times.

## IV. RESULTS AND DISCUSSION

### A. Pulse attenuation in the chain with mass ratio 0.98

In experiments short pulse was generated by the impact of a spherical steel particle, the same as the spherical particles in the chain. Impactor mass was close to the half of cell mass ($m_s + m_c$). The drop height of the striker is kept equal to 3 cm resulting in an amplitude of the reference signals (around 110 N) recorded by the gauge embedded into the 4th cylinder.

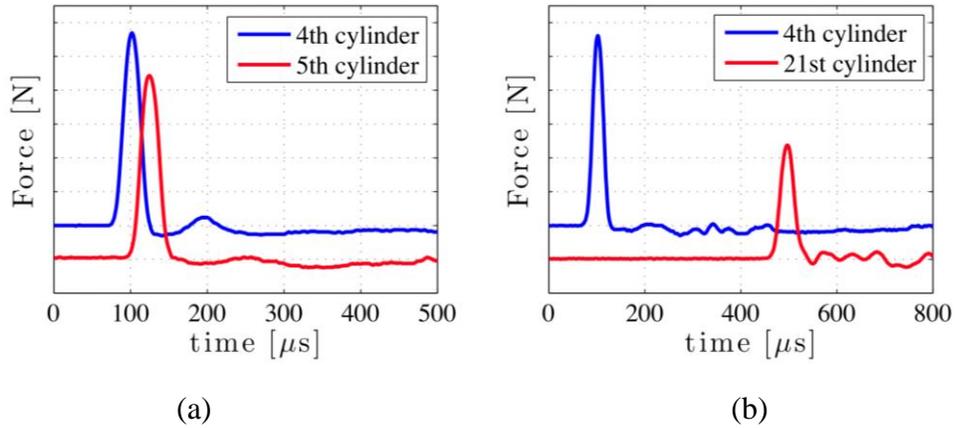

FIG. 2. (Color online) Experimental results. Single pulse propagating through a chain with the mass ratio of 0.98. (a) Signals correspond to sensors embedded into the 4$^{th}$ and 5$^{th}$ cylinder. (b) Signals correspond to sensors embedded into the 4$^{th}$ and 21$^{st}$ cylinder. Pulses were excited by the impact of a spherical steel particle. The vertical scale in both figures is 20 N and zero time is arbitrary, the curves are offset for visual clarity.

The impact resulted in a single, solitary like pulse (Fig. 2). The pulse speed (618±19 m/s) was calculated using distance between sensors and time interval between maximums of bell shape signals using the data from the sensors embedded into 4th and 5th cylinders.

A reference pulse duration was equal to 55 μs based on records from the sensor embedded into the forth cylinder. Based on these data the length of the pulse was equal to 2.5 cell sizes (cell is composed from sphere and cylinder), which is close to the expected solitary wave length in the chain with equal masses (5 particles).

As the signal propagates through the system, the amplitude decreases and the time width of the pulse increases which is clearly seen by comparison of signals corresponding to the sensors in $4^{th}$ and $21^{st}$ cylinders presented in Fig. 2(b). The pulse speed (611 ± 6 m/s) calculated based on records of gauges embedded into $15^{th}$ and $21^{st}$ cylinders was slightly lower than a speed measured in the interval between $4^{th}$ and $5^{th}$ cylinders, due to the attenuation of the pulse amplitude (Fig. 2). The shape of the pulse detected by the gauge in the $21^{st}$ cylinder was slightly nonsymmetrical probably due to dissipation. The length of this pulse (based on its speed 611 m/s and duration) was equal to 3.5 cell size, being larger than the width of the reference pulse detected by the sensor embedded into the $4^{th}$ cylinder.

It is interesting to compare the speeds and width of the pulses with exact analytical solution for the nondissipative chain obtained in long wave approximation for the chain without static precompression. This is possible to do because the amplitude of the pulse in the wave is much larger than gravitational precompression. In numerical calculations we demonstrated that the pulse shape and speed at the investigated distances from entrance were negligibly affected by gravitation.

The separate numerical calculations with mass ratio equal 1 were performed and demonstrated insignificant difference of properties of solitary waves under the same impact at investigated number of particles in comparison with the case when mass ratio was 0.98. For example, in the presence of gravitation a difference in amplitudes at 43th particle was 0.02% in the chains with mass ratios 1 and 0.98.

If we neglect the difference between masses of spheres and cylinders, then the speed of the solitary wave ($V_s$) can be calculated using the following equation connecting parameters of solitary wave solution in a long wave approximation [5]

$$V_s = \frac{2}{\sqrt{5}} c \xi_m^{1/4} = \left(\frac{16}{25}\right)^{1/5} c^{4/5} v_m^{1/5}, \tag{13}$$

where $\xi_m$ is the maximum strain equal to $2\delta_m/(h+2R)$, $\delta_m$ is the maximum change of the distance between centers of neighboring sphere and cylinder, $v_m$ is a maximum particle velocity in a solitary wave and the constant $c$ corresponds to non-dissipative contact of the sphere and plate

$$c^2 = \frac{E}{6(1-v^2)m} \sqrt{\frac{R}{2}} (h+2R)^{5/2}. \tag{14}$$

In experiments we measure force acting on the gauge embedded inside the cylinders. The relation between the maximum of this force and speed of solitary wave can be satisfactory described by maximum force acting between neighboring particles (Eq. 15), similar to [5] with constants adjusted for the array of cylinders with flat sides and spheres

$$V_s \approx \frac{h+2R}{\sqrt{5}} \left(\frac{2ER^{1/2}}{3m^{3/2}(1-v^2)}\right)^{1/3} F_m^{1/6}. \tag{15}$$

This equation demonstrates that the speed of the solitary wave depends on the cell size ($h+2R$). It can be different for the chains with the same masses of spherical particles and cylinders, if the cylinders have different heights $h$. It should be mentioned that relation between speed of solitary wave and maximum force acting between particles (Eq. 15) uses only the leading approximation in a Taylor series for relative displacements between neighboring particles in the discrete chain and strains in the continuum limit.

The pulse speed calculated using Eq. 15 and the value of the maximum force (108 N), recorded by the gauge embedded in the 5$^{th}$ cylinder (Fig. 2(a)), is 593±12 m/s, which is close to the pulse speed (618±19 m/s) in the experiments. The speed of the pulse (547±12 m/s), estimated using Eq. 15, based on the average force amplitude (68 N) detected by the sensors in the 15$^{th}$ and 21$^{st}$ cylinders was lower than the average speed (611 ± 6 m/s) measured in the interval between 15$^{th}$ and 21$^{st}$ cylinders (based on the distance and corresponding time interval). It should be mentioned that $F_m$ in the Eq. 15 is the maximum force acting on the contact between particles, but sensors are embedded inside particles, recording the average force on the corresponding contacts [5, 39]. This average force for the solitary wave stress pulse at low precompression is about 1.4 times smaller than the maximum force [39], contributing to the difference in speeds calculated from the maximum of recorded force and time intervals. We can conclude that despite the wider pulse length and its asymmetric shape its speed was close to the speed of the solitary wave in the uncompressed uniform chain despite dissipation present in the system. Thus this system of cylinder and spheres supports attenuated localized stress pulses close to the solitary like waves predicted in a long wave approximation in agreement with previous observations in the chains composed from spheres only.

Numerical calculations of the pulse propagating in a discrete chain with mass ratio 0.98 were carried out for the same system parameters. The main focus of our research was on pulse propagation inside the system. To make appropriate comparison with experiments (impact velocity 1.2 m/s), in numerical calculations the velocity of impactor (1.06 m/s) was adjusted to reproduce the same amplitude and shape of the reference pulse in $4^{th}$ cylinder as in experiments and trace the evolution of this pulse as it propagates inside the chain. Fig. 3 shows the comparison between the numerical calculations (with and without dissipation) and the experimental results. It is worth mentioning that gravitational force is considered in our numerical calculations. However, since this is a relatively short chain, calculations without gravitational force show similar results with largest difference in amplitudes of waves about 5% for longest travelling distances investigated in the paper.

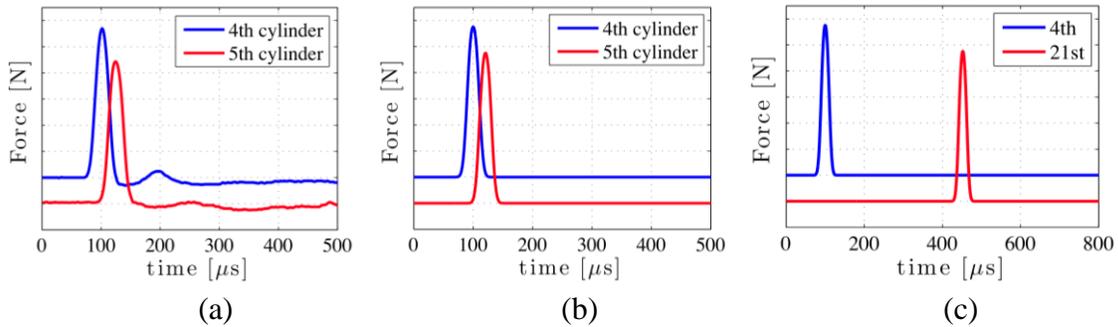

FIG. 3. (Color online) Comparison of experimental results and numerical calculations of the pulse propagating through the system which has a mass ratio of 0.98. Sensors are placed in the 4th, 5th and 21st cylinders. (a) Experimental results, (b), (c) numerical calculations (without dissipation) related to the forces in the corresponding cylinders. The vertical scale is 20 N and the curves are offset for visual clarity and zero time is arbitrary.

In numerical calculations without dissipation the solitary wave was quickly formed at the first few particles. It propagated without noticeable changes in its speed (648 m/s) and shape with value of particle velocity in the maximum being equal to 0.74 m/s. Speed of solitary wave with the same amplitude of particle velocity using analytical solution (Eq. 13) was equal to the same value 648 m/s. Their equal values are in the agreement with the comparison of solitary wave speeds in discrete chain and in analytical solution in continuum limit at the same amplitude of particle velocities, the difference being less than 1% [5]. At the same time the value of solitary wave speed in numerical calculations was close to the experimental value of 618±19 m/s, corresponding to the solitary wave with amplitude similar to numerical calculations. Pulse length in numerical calculations was equal to 2.5 cells similar to experimental results.

Detailed comparison between analytical solution in a long wave approximation and results of numerical simulations for a discrete chain were discussed in [6]. Assuming that total displacement in numerical calculations for discrete chain during the passage of the solitary wave is equal to displacement of particles derived from the exact solution for strain the authors found amplitude of displacement in latter solitary wave solution. With this value of strain amplitude in continuum limit the calculated speed of solitary wave was close to the speed in numerical calculations within 2%. At the same amplitude of displacement the calculated maximum of particle velocity was 12% lower than in numerical calculations of discrete chain. A calculated maximum force using two terms in a Taylor series was 13% lower than in numerical calculations of discrete chain [6]. This correspondence of exact solution for long wave approximation with results for a discrete chain explains successful use of the former solution to describe experimental results in 1-D chains made from particles of different materials where the accuracy of force measurements is about 10% [8,9,12].

It is evident that we have attenuating solitary wave in our experiments. To explain the observed attenuation of the pulse in experiments the dissipation was modeled by introducing a viscous terms into all contact interactions, as described above (Eqs. 8-12). Fig. 4 shows the comparison between the numerical calculations with dissipation ($\mu = \mu_1 = 6$ kg/s) and the experimental results.

In numerical calculations pulse at $21^{st}$ cylinder has a tail with small constant positive amplitude equal to 3 N (see insert to Fig. 4(b)), which is in qualitative agreement with the modification of corresponding signal shape in experiments (Fig. 4a). This is also in agreement with the influence of viscous dissipation on the shape of short pulses which was investigated in [35, 36].

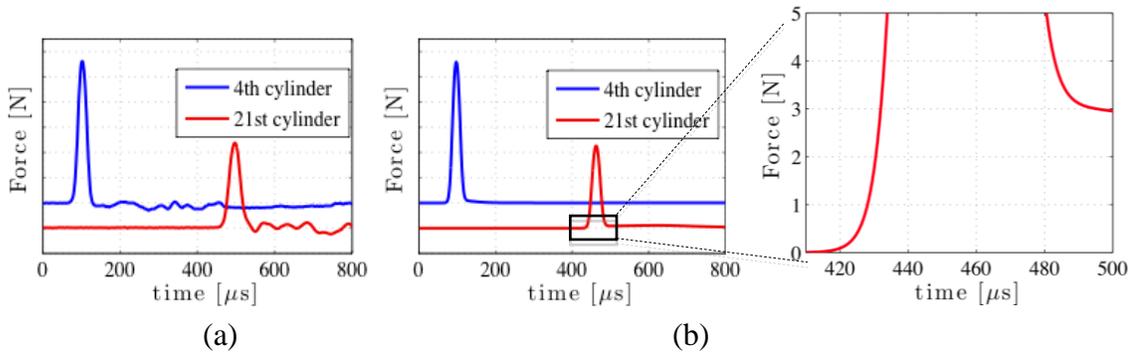

FIG. 4. (Color online) Experimental results and numerical calculations of the pulse propagating through the system with mass ratio of 0.98. (a) Experimental results, sensors are placed in the $4^{th}$ and $21^{st}$ cylinders, (b) Numerical calculation with damping coefficient 6 kg/s related to the forces in the corresponding cylinders. The vertical scale in (a), (b) is 20 N. The curves are offset for visual clarity and zero time is arbitrary.

Comparison of attenuation of the relative pulse amplitude ($A_i/A_4$) in experiments and numerical calculations using damping coefficients 0, 4 and 6 kg/s is shown in the Figs. 5. It is clear that introduction of the viscous dissipation correctly explains the signal amplitude decay. Both damping coefficients satisfactory describe the experimental data, and damping coefficient 6 kg/s provides a better fit at largest investigated distances from the impacted end.

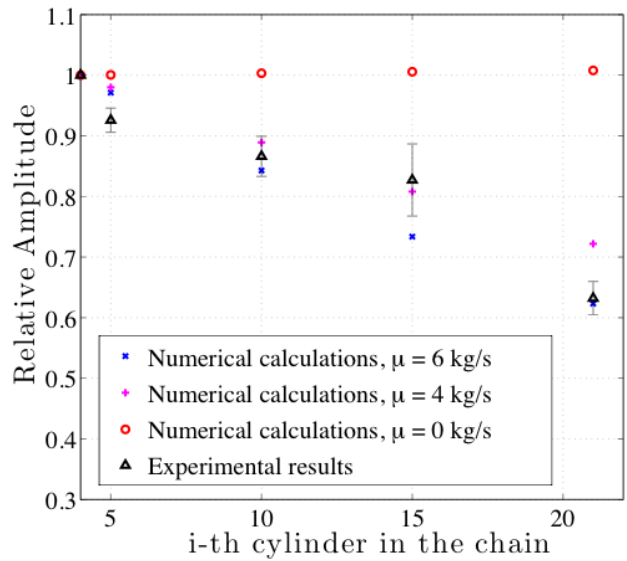

FIG. 5. (Color online) Relative amplitude of signal at i-th cylinders (corresponding to different depths) with respect to the amplitude of the reference pulse detected by the sensor in the 4th cylinder in the chain with mass ratio 0.98, experimental data and numerical results with different damping coefficients.

The decay of relative pulse amplitude due to damping as it travels through the chain in numerical calculations (damping factor equal 6 kg/s) and in experiments is satisfactory described by exponential function with value of exponent equal to 0.028 (Fig.

6). Exponential decay in numerical calculations due to plastic deformation on the contacts was observed in the paper [23].

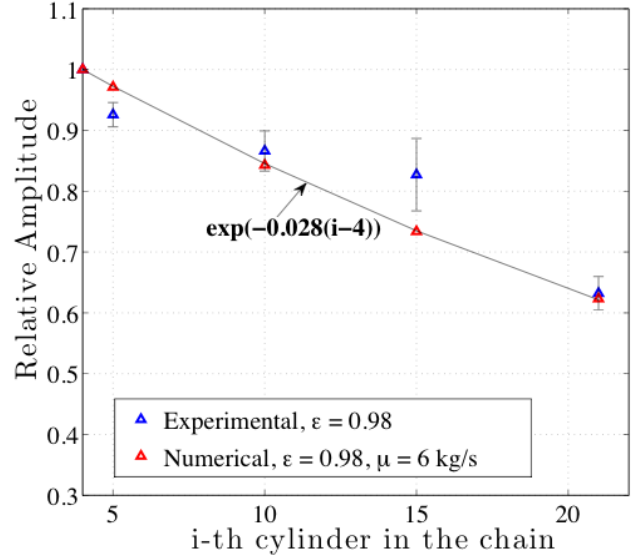

FIG. 6. (Color online) Attenuation in experiments (blue) and in numerical calculations with damping coefficient 6 kg/s (red) comparing with exponential decay.

The dissipation in one mass chain resulted not only in attenuation of the amplitude of solitary like pulse, but also in the tail wave following this pulse, first introduced in [35, 36]. No such tail was detected in nondissipative chain. This small amplitude tail at relatively close distance from the impacted end is clearly identified in numerical calculations shown in Fig. 7. At larger distance from the impacted end the clear separation of the leading solitary wave and shock like pulse are observed in numerical calculations (Fig. 7).

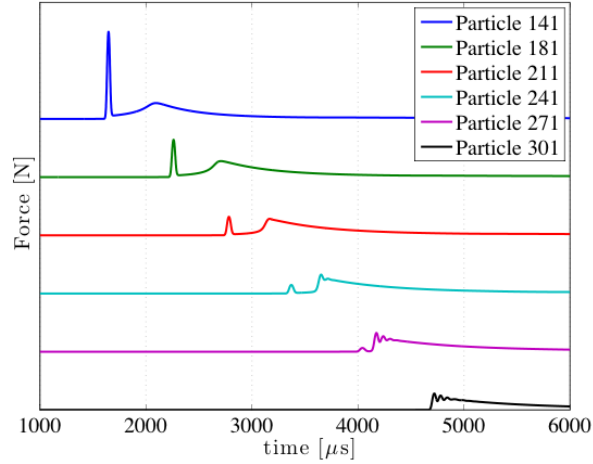

FIG. 7. (Color online) The formation of two wave structure (solitary wave followed by shock wave) in one mass chain impacted by sphere with velocity 1.45 m/s, damping coefficient 6 kg/s. The *y*-axes scale for all curves representing forces in corresponding particles (all of them are cylinders) are offset by 10 N for clarity.

The mechanism of this two waves pattern was provided in [35, 36]. The faster attenuation of leading solitary pulse is due to the larger gradients of particle velocity due to small space scale of this pulse (about 5 particles). When its amplitude becomes smaller than amplitude of following shock wave they start a process of convergence resulting in oscillating shock wave. This unique process of two wave structure generated by dissipation was also observed at a larger damping coefficient 10 kg/s and 15 kg/s. For example, in the latter case two waves were formed at the vicinity of 60th cylinder and they converged approaching the 70th cylinder. Only oscillatory shock wave remained at the depth corresponding to the 80$^{th}$ cylinder. At damping coefficient 100 kg/s two waves pattern was not formed, instead monotonous attenuating shock wave was observed.

B. Pulse attenuation in the dimer chain with mass ratio 0.55

Pulse in the dimer chain with mass ratio 0.55 (mass of sphere equal 2.085 g and mass of cylinder was increased to 3.77 g) in experiments was excited by the same impactor (mass 2.085 g) at the same velocity as in the previous chain with mass ratio 0.98. This allows comparison of pulse transformation in both chains under identical conditions of impact and contact interaction between particles. It is known that the sphere/sphere chain with mass ratio 0.55 does not support stationary solitary waves [20,21].

In this paper a different cylinder/sphere dimer chain is investigated. The results of numerical calculations for transmission of dynamic force in these chains with fixed contacts depending on mass ratio of particles for a chain composed from 42 particles (21 spheres and 21 cylinders) are shown in Figure 8. It is clear that in nondissipative chains the value of mass ratio corresponding to the position of global minimum is close to the values in non-fixed contacts sphere/sphere chains [20,21].

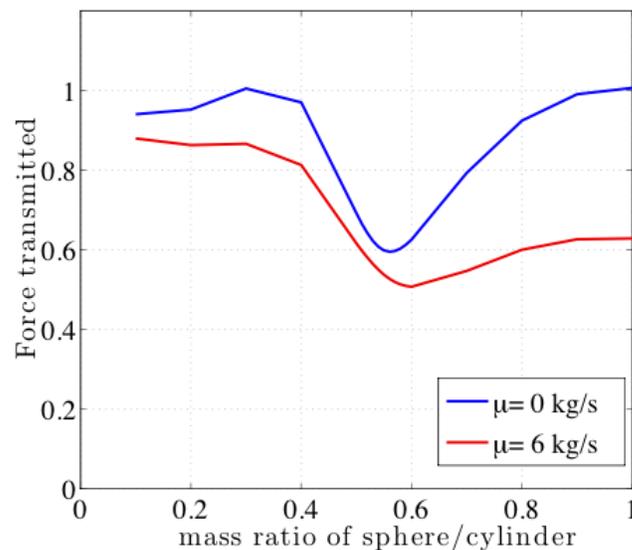

FIG. 8. (Color online) Dependence of the force in the 21st cylinder normalized with respect to the force at the entrance (at 4$^{th}$ cylinder) on mass ratio in nondissipative and dissipative chains with fixed contacts. Mass of sphere was kept the same and mass of cylinder changed keeping fixed contacts.

The numerical calculations with damping coefficient 6 kg/s, introduced to explain experimental results with the chain having mass ratio 0.98 and the same contacts as dimer chain are presented in Fig. 8 also. It is clear that dissipation shifts the position of the global minimum toward larger values of mass ratio (0.6) in comparison with nondissipative chain. It should be emphasized that the change of transmitted force in dissipative chains is not symmetric with respect to global minimum - significantly larger reduction of amplitude is observed at mass ratios in the interval 0.6-1 than in the interval 0.1-0.6. We explain this nonsymmetric behavior by larger gradients of particle velocity between neighboring particles in chains with smaller mass differences. This mechanism is qualitatively similar to the difference in attenuation between chains with mass ratios 0.98 and 0.55 explained later.

In agreement with this prediction (Fig. 8), impact by stainless steel sphere with velocity 1.4 m/s did not generate a single solitary wave in experiments (Fig. 9), unlike in previous case with practically equal masses of sphere and cylinders (Fig. 2). Instead a leading pulse was followed by series of smaller amplitude pulses. The leading pulses captured by the gauge imbedded in the heavier particles (cylinders) have double peaks repeatable in all experiments. In numerical calculations of the chain with mass ratio 0.5 leading pulses of particle velocity for light particles also had two peaks, followed by oscillating velocity profiles with negative velocities at some moments (Fig. 1.18(b) in [5]). As the pulse propagates inside the chain it attenuates and the amplitude difference between these two peaks becomes smaller (Fig. 9 (b)). The pulses following the leading doable peak transformed into oscillatory tail later (compare signals from sensors embedded in 4th and 5th cylinders with signal from the sensor in 21st cylinder). The decrease in the pulse amplitude is caused by the fact that this chain does not support solitary waves and dissipation.

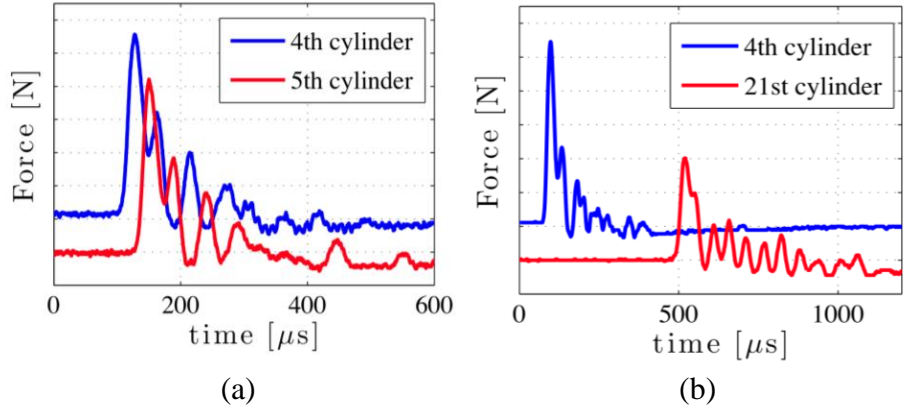

(a)            (b)

FIG. 9. (Color online) Experimental results. Stress pulses propagating through a chain with the mass ratio 0.55, mass of cyliders is larger than mass of spheres. (a) Signals correspond to sensors embedded into the 4th and 5th cylinder. (b) Signals correspond to sensors embedded into the 4th and 21st cylinder. Pulses were excited by the impact of a spherical bead the same as the spherical particles in the chain. The vertical scale in both figures is 20 N and zero time is arbitrary, the curves are offset for visual clarity.

The speed of the leading pulse based on the experimental data recorded by the gauges installed at 4th and 5th cylinder is 771 m/s. The smaller pulse speed of 712 m/s was calculated based on gauges installed in 4th cylinder to 21st cylinder. The 10% decreasing in the pulse speed is due to the attenuating pulse amplitude in the interval between 4th and 21st cylinders.

In numerical calculations the force acting on the gauges imbedded into the corresponding cylinders were found by averaging the forces acting at their contacts with neighboring spheres similar to [39]. Figures 10 (b),(c) present results of numerical modeling of nondissipative chain related to the experimental data (Fig. 10(a)). In numerical calculations reference pulses (corresponding to the force in 4$^{th}$ cylinder) with amplitude similar to experiments were generated using a lower impactor velocity (1.3 m/s) accounting for the dissipation of the signal prior its arrival on 4$^{th}$ cylinder in experiments (impactor velocity 1.4 m/s).

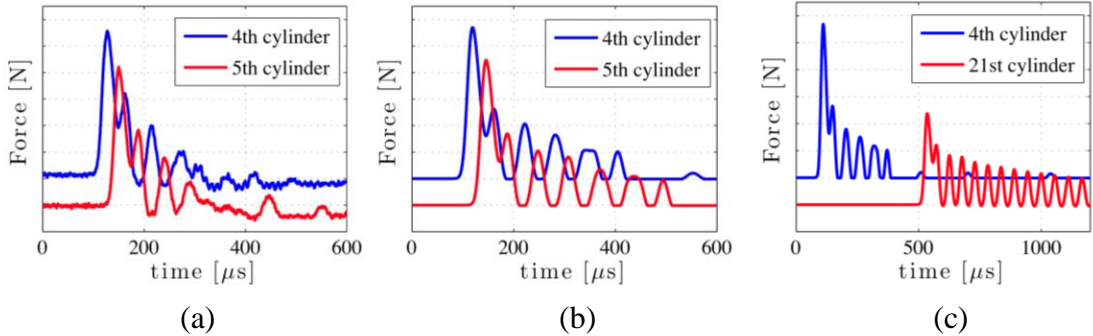

(a)                  (b)                  (c)

FIG. 10. (Color online) Comparison of the experimental results and numerical calculations (without dissipation) of the pulse propagating through the dimer system with mass ratio 0.55, mass of cylinders is larger than mass of spheres. (a) Experimental results; (b), (c) numerical calculations (without dissipation) related to the forces in corresponding cylinders. The vertical scale is 20 N and the curves are offset for visual clarity and zero time is arbitrary.

The pulses in the numerical calculations in nondissipative chain ($\mu = 0$ kg/s) have the double peak feature, and the leading pulse is followed by a similar number of pulses as in experiments. Only small decrease of pulse amplitude (when it propagates from 4$^{th}$ to 5th cylinder) from 111 N to 104 N, and from 111 N to 107 N was observed in experiments and numerical calculations, correspondingly. In numerical calculations pulses, trailing the main double peak pulse, are clearly separated from each other (Fig. 10(b) and (c)), unlike in experiments where their separation is less evident (Figs. 9 and 10(a)). In experiments only second pulse is clearly separated from the leading one in the 4th and 5th cylinders. In numerical calculations, the speed of the leading pulse traveled from 4th to 5th cylinder is 733 m/s. This speed is close to the experimental result (771 m/s). The pulses frequency spectrums are similar in experiments and numerical calculations.

The number of secondary pulses steadily increases as the wave propagates deeper into the dimer system in experiments and in numerical calculations (Figs. 9 and 10). This is an important specific feature for a two mass system with investigated mass ratio because it provides a nondissipative mechanism of pulse decay due to energy leak from the leading pulse into increasing number of secondary pulses in oscillatory tails. The mechanism of formation of these secondary pulses without gravitational loading was explained in [20]: for a general value of mass ratio (except some specific values) typically the light bead loses contact with its left neighboring heavy bead retaining a small portion of the energy of the propagating pulse and generating traveling waves in oscillating tails.

It should be mentioned that in numerical calculations the shape of the leading pulses and their amplitudes were negligibly affected by the gravitation in the investigated chains composed from up to 50 cylinders.

At the same time the difference in behavior between gravitationally loaded and free of precompression chains may be very significant for longer chains. For example, in nondissipative, noncompressed chains at relatively short distances from the impacted end the leading double peak, corresponding to force in the heavy particles (cylinder), is followed by a regular sequence of localized pulses. Force in spheres (light particle) has a single leading peak, followed by periodic sequence of double peaks pulses. These regular patterns were transformed correspondingly into two clearly separated triple peaks leading pulses in 499th particle (cylinder) and double peaks in 500th particle (sphere), followed by the chaotically oscillating trails in both cases.

In the same chains under gravitation loading (also with mass ratio 0.55) only one leading double peak was observed up to the distance about 400 particles. At larger distances the sequence of single peak pulses started to form with slowly attenuating leading pulse (composed from 15 particles) clearly separated from the rest at the 900th particle.

The difference between experiments and results of nondissipative numerical calculations suggests introduction of dissipative damping as with chain with similar masses. By design both chains have identical contacts. Thus the dissipative damping should be similar to previous case of the chain with mass ratio 0.98. It should be mentioned that the dimer systems with different mass ratio, when composed from spherical particles [20,21], could experience a different damping on the non-fixed sphere/sphere contacts than in our systems with fixed plane/sphere contacts.

Fig. 11 shows comparison of numerical results (with different damping coefficients) with the experimental data.

The decrease in the first peak amplitudes of the leading pulse at $21^{st}$ cylinder in the numerical calculation from 109 N to 59 N (with damping coefficient 4kg/s) and from 109 N to 57 N (with damping coefficient 6 kg/s) are similar to the experimental results (109 N to 60 N).

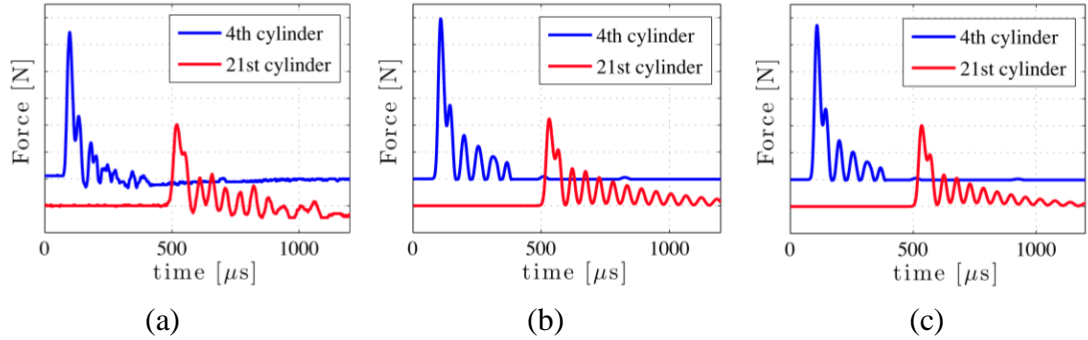

(a)                      (b)                      (c)

FIG. 11. (Color online) Results of experiments and numerical calculations of the pulse propagating through the dimer system, mass of cylinders is larger than mass of spheres. (a) experimental results (sensors are placed in the 4th and 21st cylinders), (b) and (c) results of numerical calculations related to the forces in $4^{th}$ and $21^{st}$ cylinders with $\mu = 4$ kg/s and $\mu = 6$ kg/s, correspondingly. The vertical scale is 20 N and the curves are offset for visual clarity, zero time is arbitrary.

As the signal propagates through the dissipative system, the amplitudes of the leading double peak pulse decrease, while the number of following pulses increased in numerical calculations in agreement with experiments (Fig. 11).

Figure 12 presents experimental and numerical results with various damping coefficients related to the amplitude attenuation of leading pulses with cylinder numbers. It is clear that in numerical calculations the leading pulse in the chain with mass ratio 0.55 is attenuating even without dissipation. This is due to the fact that the investigated system with mass ratio 0.55 does not support stationary solitary waves. Without dissipation, decrease of relative amplitude of the force at $21^{st}$ cylinder is about 40% and this decay is caused solely by dispersion. Dispersion in the pure nonlinear system means dependence of a wave speed on wave length caused by mesostructure, e.g., size of the particles. Like in a weakly nonlinear systems (the equation for corresponding granular chain can be found in [5], see Eq. (1.7) there) the dispersion in strongly nonlinear systems balances strong nonlinearity resulting in a strongly nonlinear solitary wave unique for pure nonlinear systems in one mass chain or in two mass chains at specific values of mass ratio [20]. In two mass chains at arbitrary mass ratio nonlinearity is nor balanced by dispersion. In this case solitary waves are not supported by a system resulting in pulse decay (being maximized at mass ratio 0.59) even in nondissipative chains [20]. The difference with weakly nonlinear system is that in strongly nonlinear case the dispersion term in corresponding wave equation is also nonlinear (Eq. 1.20-1.23 in [5]).

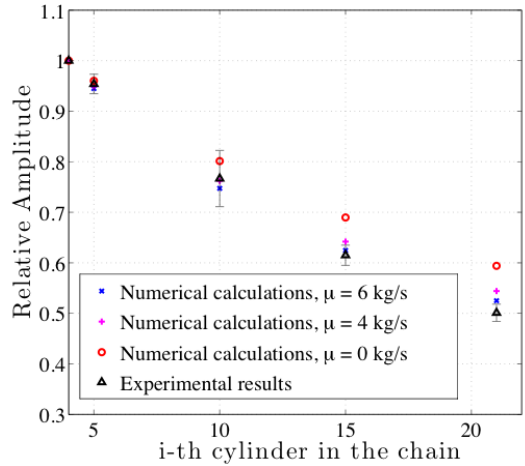

FIG. 12. (Color online) The comparison of the experimental results and data from the numerical calculations with different values of damping coefficient in the dimer chain with mass ratio 0.55. Relative amplitude of leading stress pulse at different depths (number of cylinders are shown on the horizontal axes) is calculated with respect to the amplitude of the reference pulse in the 4th cylinder. Mass of cylinders is larger than mass of spheres.

Numerical calculations in dissipative chains (with damping coefficients 4 kg/s and 6 kg/s) demonstrate larger attenuation with decrease of amplitude about 50%. The similar attenuation was observed in experiments. The role of dissipation is significant starting at 15th cylinder, at smaller distances its role is negligible and signal attenuation is mostly due to dispersive effects (Fig.12).

Based on the attenuation of the amplitudes at different positions (Fig.12) inside the chain we conclude that numerical calculations with damping coefficients 6 kg/s satisfactory fit experimental data. This value also provided the best fit for experimental data in chain with mass ratio 0.98. It is explained by the same nature of contact interaction between surface of the steel cylinder and sphere of the same diameter causing the same mechanism of dissipation.

The amplitude of the pulse decays exponentially in experiments and in numerical calculations as it travels through the two mass chain due to the combination of dispersive and dissipative mechanisms. As a result in the two mass chain the exponent (0.042) is larger than in the case of mass ratio equal to 0.98.

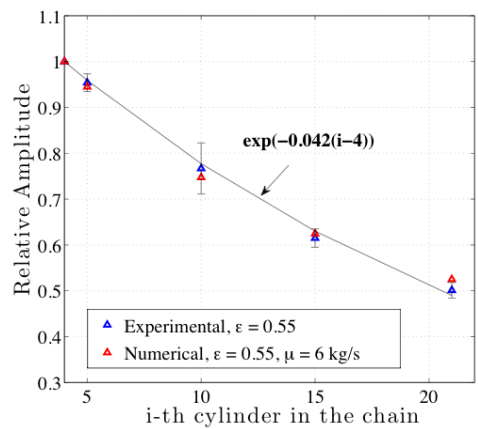

FIG. 13. (Color online) Attenuation of relative amplitude in experiments (blue) and in numerical calculations with damping coefficient 6 kg/s (red) comparing with exponential decay. Relative amplitude is calculated with respect to the amplitude of the reference pulse in the 4th cylinder. Mass of cylinders is larger than mass of spheres.

The qualitative difference between two mass nondissipative and dissipative chains is the opening of gaps on both sides of 21st cylinder in the former chain as demonstarted in Fig. 14.

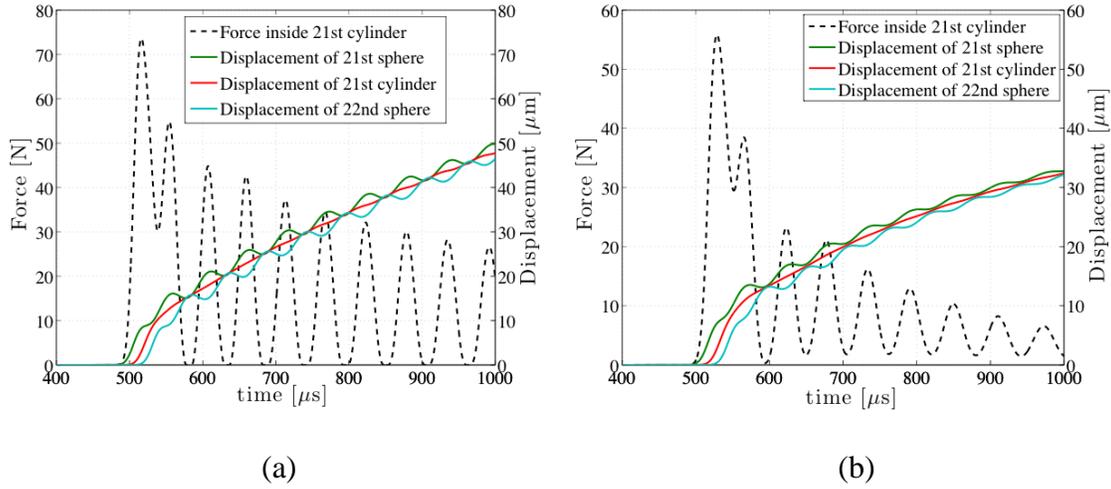

(a)          (b)

FIG. 14. (Color online) Comparison of the forces inside $21^{st}$ cylinder and out of phase displacements of neighboring spheres in numerical calculations: (a) damping coefficient equal to zero, opening of gaps on both sides of $21^{st}$ cylinder is evident at moments corresponding to zero forces and (b) no openings of gaps at damping coefficient equal to 6 kg/s. In both cases chain with mass ratio 0.55 impacted by sphere with velocity 1.45 m/s.

In dissipative chains at damping coefficient 6 kg/s there are no gaps open (Fig. 14(b)). In both cases the oscillatory force profile in 21st cylinder is observed due to oscillating motion of spheres clearly seen in their displacements curves. Increase of damping coefficient to 10 and 15 kg/s reduces amplitudes of oscillations of spheres also without gaps openings and transforming the sequence of separated pulses into oscillatory tail.

Shape of the wave during the propagation of these pulses into the larger depths (beyond 90th cylinder) changed into attenuating triangular shock like oscillatory pulse with decreasing amplitudes of oscillations with incerse of damping coefficient to 10 and 15 kg/s.

The similar impact on dimer chains with damping coefficient 100 kg/s resulted in fast attenuating, smooth nonsymmetric dispersive pulses with increasing duration and ramp time being much shorter than the tail. At this damping coefficient (100 kg/s) shock waves in one mass chain and in two mass chains were practically identical at similar particle numbers with slight differences in speed propagation and amplitudes.

Two wave structure of the attenuating pulse (leading solitary like pulse clearly separated from the following shock like wave) observed at some range of distances from the impacted end in the chain of equal masses at damping coefficient 6 kg/s (Fig.7) was not detected in dimer chains.

It is interesting to compare relative attenuation in a system with mass ratio 0.55 to the system with mass ratio 0.98 (Fig. 15) at the different depths, but corresponding to the same number of cells (dissipative contacts) for chain with different damping coefficients. Though the damping coefficient 6 kg/s describes experimental data satisfactory for both chains (Figs. 5 and 12), it is interesting to compare systems with different mass ratio at larger damping coefficients which may correspond to the larger plastic deformation at the contacts or to the chains immersed in liquid. The difference can be expected because solitary wave is not supported in the former case (thus localized pulse is attenuated (Fig. 15(a)) even without dissipation losing energy into the oscillatory tail). At the same time attenuation due to dissipation might mask this difference if widths of pulses are different and contact dissipation is the main source of attenuation. The results are presented in Fig. 15 (b), (c), and (d) for damping coefficients 10 kg/s and 15 kg/s, and 100 kg/s, correspondingly.

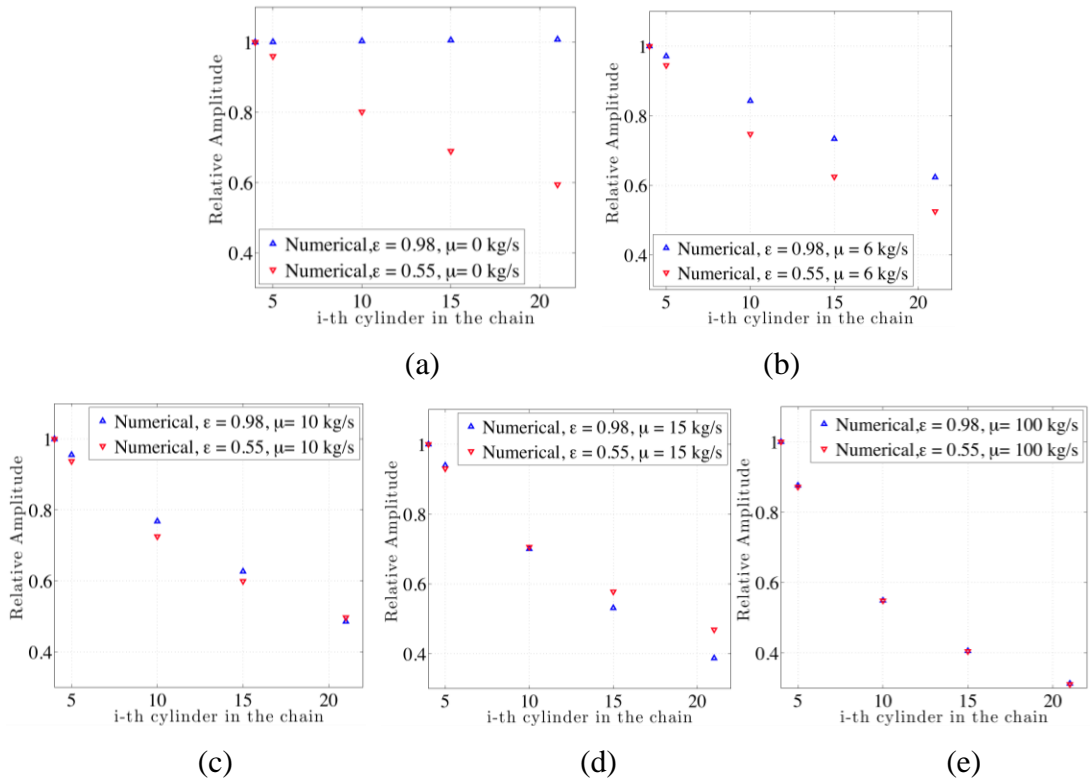

FIG. 15. (Color online) The attenuation of the relative pulse amplitude (with respect to the amplitude of the reference pulse in the 4th cylinder, mass ratios 0.98 and 0.55) and change of decay efficiency in two systems with increased damping coefficient at the same contact number in the chains at corresponding values of damping coefficients 0 (a), 6 kg/s (a), 10 kg/s (b), 15 kg/s (c), and 100 kg/s (d). Mass of cylinders is larger than mass of spheres (2.085 g).

For systems with damping coefficient 6 kg/s (Fig. 15 (b)), which gives the results most close to experimental data, the relative amplitude decrease in the system with mass ratio 0.98 (where dissipation is the only mechanism for attenuation) at $21^{st}$ cylinder contact is about 40%. The amplitude decrease in a system with mass ratio 0.55 at the same contact number is higher being equal close to 50%. The 10% difference is due to the presence of nonlinear dispersion (not balanced by strong nonlinearity) in the latter system.

With the increase of damping coefficient to 10 kg/s the amplitude decrease with increase of contact number in both systems is very close (Fig. 15(c)), manifesting that nonlinear dispersion is not a major factor in decay in the chain with mass ratio 0.55.

Further increase of the damping coefficient to 15 kg/s demonstrates the surprising result - the attenuation in 0.98 mass ratio system becomes larger than in the two mass system with optimal mass ratio 0.55 (Fig. 15(d)), which provides maximum decay in nondissipative dimer chain (Fig. 8). We explain this phenomenon by the difference in shapes of particle velocity profiles in investigated systems (Fig. 16). From this figure (Fig. 16(d)) it is clear that dissipation at the damping coefficient 15 kg/s does not result in significant differences of pulses space scales in comparison with the case without dissipation, only adding tails in particle velocities which amplitude is increasing with increase of damping coefficient. This demonstrates that strong nonlinearity and nonlinear dispersion caused by periodic mesostructure control the shape of the pulses in both systems even in the presence of dissipation.

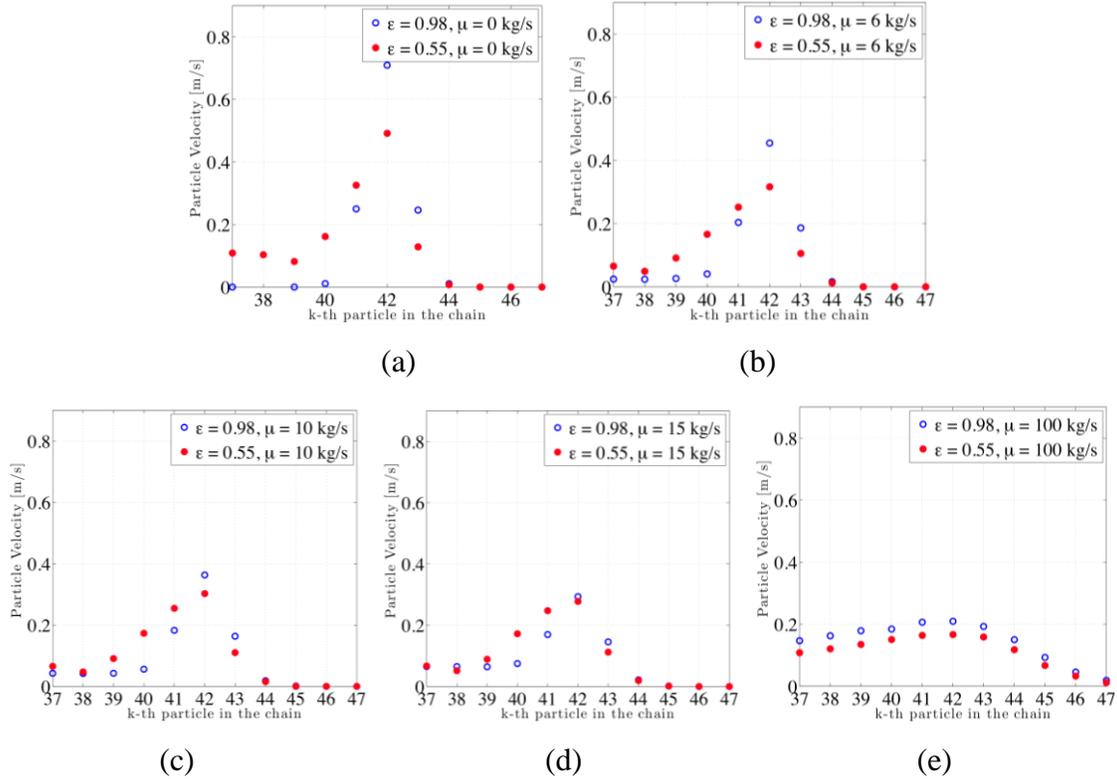

FIG. 16. (Color online) The change of particle velocities profiles in both sytems with increase of damping coefficient. The maximum of pulse amplitude corresponds to 42nd particle (21st cylinder), odd numbers are related to spheres and even to cylinders. Numerical calculations with damping coefficients 0 (a), 6 kg/s (b), 10 kg/s (c), 15 kg/s (d) and 100 kg/s (e). Particle velocities in system with mass ratio 0.98 (blue, open circles) and particle velocities in system with mass ratio 0.55 (red, solid circles). Mass of cylinders is larger than mass of spheres (2.085 g).

The important difference between chains with mass ratio 0.98 and 0.55 is that the former chain supports a narrow solitary wave composed from only 5 particles, with major gradients of velocity just between two particles at the front and two particles at the back of this wave [5]. This large differences between velocities of neighboring particles in one mass chain result in larger viscous dissipative losses (Eqs. 8-12) in comparison with two mass chain were localized pulses (not solitary waves) have a longer dimensions and thus a smaller gradients of velocity between neighboring particles. This difference may explain the reversal in impact mitigation effectiveness of these systems with increased damping coefficients.

Moreover two mass chains in the discussed case (heavy cylinders/light spheres) would be heavier at the same number of particles than one mass chain (light cylinders/light spheres). Thus at this level of dissipation (15 kg/s) one mass chain with the same number of particles has a smaller mass also making it a better protector against impact pulse with a short duration.

From comparison of pulse attenuation at damping coefficients 10 kg/s and 15 kg/s we can conclude that the former value is close to critical value corresponding to the reverse of performance of these systems with respect to pulse amplitude decay. This transition corresponds to the prevailing influence of dissipation over decay caused by mesostructure in dimer chain.

The further increase of damping coefficient to 100 kg/s makes the pulse shape in both system very similar (nonsymmetric triangular pulse, Fig. 16(e)) resulting in negligible role of nonlinear dispersion and in practically identical attenuation in both systems (Fig. 15(e)) after pulse travelling through the same number of particles.

However, the systems with the same numbers of particles, but with different mass ratios have different total masses. For design purpose (for example if mass of the protection layer is the main design parameter, e.g., in helmets), it is interesting to compare the attenuation of the pulses at different values of damping coefficient in the same systems with different mass ratios after propagation not through the same number of particles (as in Fig. 15), but through the chains with the same masses. The corresponding data from experiments and numerical calculations are presented in Fig. 17, linear approximation was used to approximate the data in Fig. 17(a) based on data from Fig. 5 and Fig. 12.

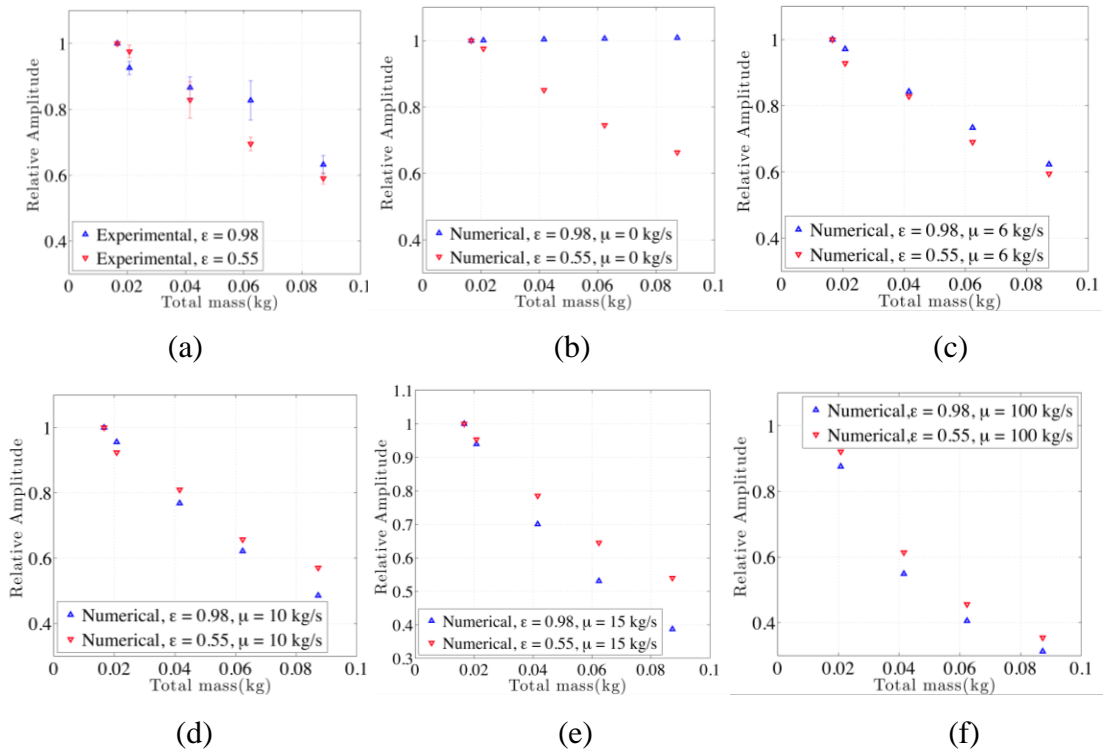

FIG. 17. (Color online) Relative amplitude (with respect to the amplitude of the reference pulse at the 4th cylinder) after propagation through the chain with the same mass at different values of damping coefficient in experiments (a) and in numerical calculations with different damping coefficients: 0 (b), 6 kg/s (c), 10 kg/s (d), 15 kg/s (e) and 100 kg/s (f). Mass of cylinders is larger than mass of spheres (2.085 g).

In experiments the attenuation of signals traveling through the length of chains having the same masses (but with different mass ratios of particles 0.98 and 0.55 composed from the same spheres (2.085 g) and cylinders having masses similar to the sphere mass and heavier, correspondingly) are quite similar (Fig. 17(a)). In numerical calculations without dissipation (chain with mass ratio 0.98) no attenuation of the pulse was observed in contrast to significant attenuation in the nondissipative chain with mass ratio 0.55, (Fig. 17(b)). The similar attenuation in experiments (Fig. 17(a)) in both chains of equal masses, despite the two mechanism of decay existing in two mass chain (dispersion and dissipation) versus only one mechanism in one mass chain (dissipation), is apparently due to the stronger effect of dissipation in the latter system.

In numerical calculations with a damping coefficient of 6 kg/s, the respective relative amplitudes are very close to experimental values for both chains (compare Fig. 17(a) and (c)). At the increased damping coefficients starting from 10 kg/s, the system with mass ratio 0.98 demonstrates faster pulse attenuation than system with mass ratio 0.55 after the travelling through the chain with the same mass (Figs. 17(d) - (f)). This difference in enhanced by a smaller number of contacts in the system with mass ratio 0.55 at the same total mass travelled by the pulse. Though one mass chain is also preferable for pulse mitigation given the same mass of the chain at the damping coefficient 100 kg/s, the difference is smaller than at lower damping coefficients 10 and 15 kg/s. This is explained by the similar gradients of particle velocity between elements in both chains at damping coefficient 100 kg/s (Fig. 16(e)) and by a larger number of dissipative contacts in one mass chain (light cylinders/light spheres) versus two mass chains (heavy cylinders/light spheres).

Thus strongly dissipative one mass chains will be again better for the impact protection at the same total mass of particles than two mass chains with larger mass of cylinders relative to the mass of spheres.

It is also interesting to compare behavior of one mass system with two mass system having the mass ratio of particles close to optimal value 0.55 (but with reduced mass of cylinders versus spheres) at the same striker impact. The attenuation of the relative pulse amplitude depending on number of travelled contacts in chains with mass ratios 0.98 and 0.55 is presented in Fig. 18. In the latter chain a mass of cylinders is smaller than mass of spheres (2.085 g), unlike in the previous case corresponding to Fig. 15.

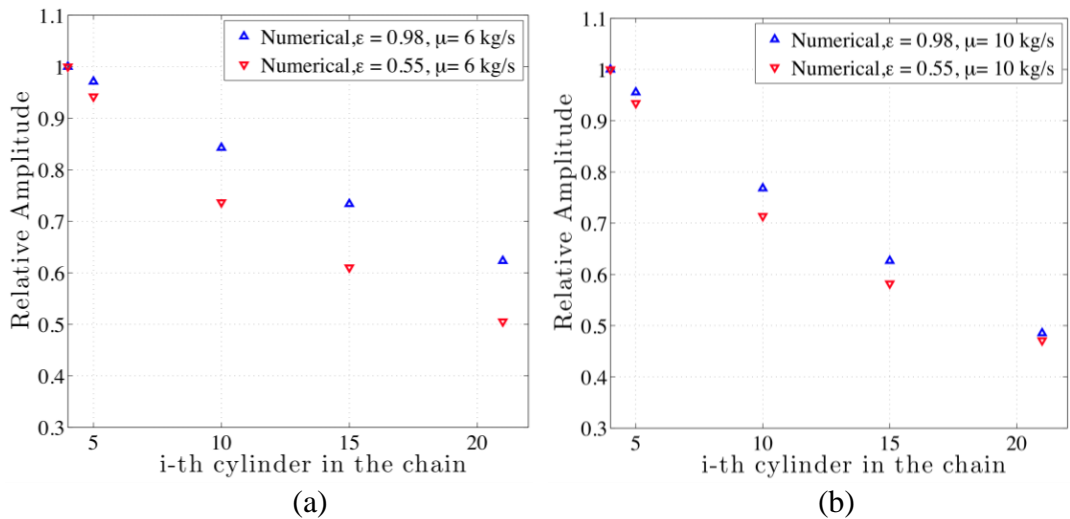

(a)　　　　　　　　　　　　　　(b)

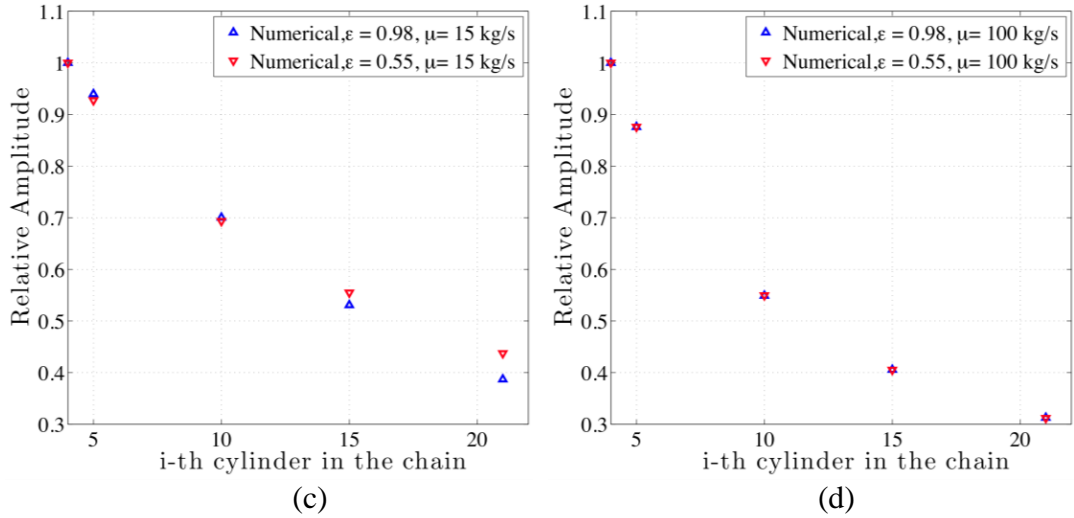

FIG. 18. (Color online) The attenuation of the relative pulse amplitude with respect to the amplitude of the reference pulse in the 4th cylinder, mass ratios 0.98 and 0.55, mass of cylinders is smaller than mass of spheres (2.085 g) and change of decay efficiency in two systems with increased damping coefficient. Relative amplitudes in numerical calculations at the same position in the chains at corresponding values of damping coefficients 6 kg/s (a), 10 kg/s (b), 15 kg/s (c), and 100 kg/s (d).

Numerical calculations (Fig. 18) show that with reduced cylinder masses and the same masses of spheres, the attenuation of relative amplitude of pulses travelled through the same number of particles in chain (number of contacts) is similar to the previous case of two mass chain with larger cylinder masses (Fig. 15). In this case also the one mass system is better in attenuating the pulse amplitude with increased damping coefficient probably for the reason that pulse in two mass system contains a larger number of particles and thus a smaller difference of particle velocities between elements. At the damping coefficient 100 kg/s the difference in attenuation is negligible when pulse travelled the same number of particles.

But in case if the optimal two mass system (mass ratio 0.55) has a smaller cylinder mass it also has a smaller total mass, at the same number of contacts, than one mass system composed from the same spheres and cylinders of equal masses. For the design purpose it is interesting to compare the attenuation of the pulses in these systems having the same mass at different values of damping coefficients. The corresponding results from numerical calculations are presented in Fig. 19.

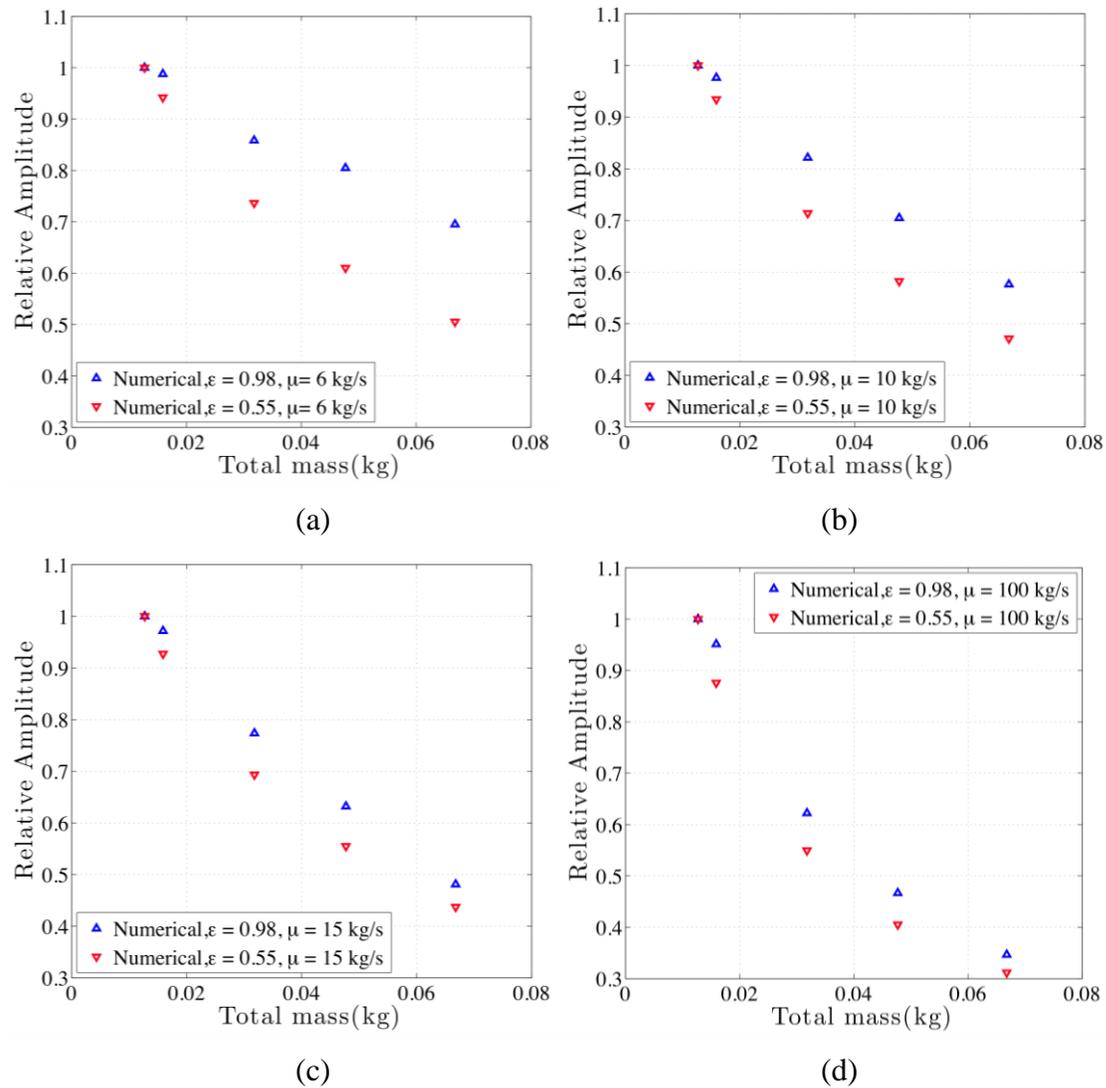

(a)

(b)

(c)

(d)

FIG. 19. (Color online) Relative amplitude (with respect to the amplitude of the reference pulse at the 4th cylinder) after propagation through the chain with the same mass at different values of damping coefficient in in numerical calculations with different damping coefficients: 6 kg/s (a), 10 kg/s (b), 15 kg/s (c) and 100 kg/s (d). Mass of cylinders is smaller than mass of spheres (2.085 g), mass ratio is 0.55.

We can see that at damping coefficient 6 kg/s (close to the value in experiments with the identical contacts) the two mass chain with mass of cylinders being smaller than mass of spheres (2.085 g), mass ratio is 0.55 is significantly better at the same chain mass. This is caused by dispersion effects in combination with smaller number of travelled dissipative contacts in one mass chain. With increased damping coefficient the difference in mitigation between thee systems becomes smaller due to the stronger effect of dissipation on signal attenuation in more narrow pulse in one mass chain explained above. Thus this chain is preferable if the mass of the attenuating system is the main design parameter, as in helmets.

## V. CONCLUSIONS

Propagation of short pulse in dimer chains with two different mass ratios (0.98 and 0.55) was investigated in experiments and numerical calculations. The same striker was used to generate short pulses in these systems to compare their effectiveness under the same impact. Both chains had the fixed cylinder-sphere contact, which keeps dissipative properties (damping coefficient) of both systems identical highlighting the role of radiation based attenuation mechanism present in the chain with mass ratio 0.55, unlike in [20,21]. The cylinder-sphere chains are more convenient with respect to design any mass ratios and keeping diameters of particles the same, which is difficult to accomplish with sphere-sphere chains placed in the channel.

The attenuation of pulse amplitude due to dissipation was modeled using linear viscosity term which qualitatively described the observed rate of attenuation in both systems at the same damping coefficient equal 6 kg/s. In both experiments and corresponding numerical calculations pulses in the system with mass ratio 0.55 attenuates faster.

The change of the dependence of the force on mass ratio in dissipative chains with fixed contacts is not symmetric with respect to global minimum. There is a relatively small change of the transmitted force for small mass ratio up to optimal ratio 0.55, but few times larger change of transmitted force was observed in the chain with small mass difference. We explain this nonsymmetric behavior by larger gradients of particle velocity between neighboring particles in chains with smaller mass differences.

Introduction of viscous damping blocks the gap opening starting at the damping ratio 6 kg/s. Thus damping not only dissipate energy, but also eliminates the process of gap openings and corresponding time scales (gap opening divided by particle velocity) characteristic for nondissipative chains.

The input into pulse decay due to strongly nonlinear dispersion effect in experiments can be illustrated by comparison between chains with the same number of identical dissipative contacts crossed by travelling pulse. In the system with mass ratio 0.98, where the only active mechanism of attenuation is dissipation, 40% amplitude decrease was observed in experiments, unlike larger decrease (50%) in the chain with mass ratio 0.55, being caused by both mechanisms of decay.

The influence of the value of damping coefficient on the relative effectiveness of these systems to mitigate identical impact was investigated numerically in the case when mass of cylinders was larger than mass of spheres. If these systems have the same number of particles or the same mass travelled by the pulse, their amplitudes in the dimer system with mass ratio 0.55 attenuate faster than in the system with mass ratio 0.98 only when damping coefficient is below some critical value (below 10 kg/s for investigated systems).

At larger damping coefficients the system with mass ratio 0.98 mitigate the same impact better than the system with mass ratio 0.55. The former chain in this highly dissipative systems has a smaller mass providing the same level of attenuation than the system with mass ratio 0.55. The one mass system is preferable system for the higher level of viscous dissipation (e.g., granular chains in liquid) because it forcefully supports the high gradients in the narrow pulse by strongly nonlinear dispersion.

At the highest investigated level of viscous dissipation (damping coefficient 100 kg/s) the pulses in both systems are of the similar width resulting in a similar viscous dissipation of pulses travelling the same number of contacts.

A different behavior of two and one mass chains with increase of damping ratio was observed with respect of shape of the propagating pulses. One mass chain excited by impact of sphere demonstrates a two wave structure (solitary wave plus oscillatory shock wave), this phenomena was not observed in two mass chain. At the largest damping ratio 100 kg/s both chains behave in a similar way.

Numerical calculations for a different system where a cylinder masses are smaller than masses of spheres show similar results to the system with larger mass of cylinders for the same number of contacts crossed by travelling pulse. But in this case dimer system with the same total mass has larger number of dissipative contacts enhancing pulse attenuation at all values of investigated damping coefficients.

These results help to select appropriate mesostructure and dissipative properties of the strongly nonlinear discrete systems based on the design limitations for protection barriers.